# A nonlinear system with weak dissipation under external force on incommensurate frequency: coexistence of attractors

Pozdnyakov M.V.[1], Savin A.V.[2], Savin D.V.[2, *]

[1] *Saratov College of Mechanical Engineering and Economics,*
*Yury Gagarin Saratov State Technical University, Saratov, Russia*
[2] *Department of Nonlinear Processes,*
*Chernyshevsky Saratov State University, Saratov, Russia*
[*]corresponding author
e-mail: savin.dmitry.v@gmail.com

The behaviour of the 2D model system – the Ikeda map – is investigated in the weakly dissipative regime under external forcing on the incommensurate frequency. Coexistence of a large number of stable invariant curves is shown. Dependence of the number of coexisting attractors on the external forcing amplitude and nonlinearity parameter is investigated.

**Introduction**

The theory of nonlinear dynamical systems studies a rich variety of phenomena distinctive for dynamical behaviour of such systems, and these phenomena can differ sufficiently depending on the class of the system under investigation. One of the main properties determining the nature of system's dynamics is the conservativity. Actually, the nonlinear dynamics of dissipative systems and of conservative ones can be divided into two branches, which in some sense developed independently for a long time (see, e.g., [1] and [2] for main phenomena of conservative and dissipative dynamics correspondingly). The most sufficient difference in dynamics of these two classes is the dynamics of the phase space volume. If the system is conservative, some initial domain taken in the phase space of a system changes its shape but conserves its volume, and hence, there exist a unique dynamical trajectory for every initial condition on the infinitely long time interval. On the contrary, if the system possesses an energy loss in the process of its evolution, the initial domain in the phase space decreases with time, and finally trajectories from all initial conditions taken in a certain domain (basin of attraction) converge to some invariant set in the phase space – an attractor. Normally a dissipative dynamical system possesses only one or few number of such attractors at certain parameter values (the latter situation is usually called multistability). It turns out that for some classes of dissipative systems the multistability can be strong enough, i.e. the number of coexisting attractors can be not very small. For example, such type of behaviour is typical for coupled systems, differential equations with delayed feedback and several other classes of dynamical systems (see [3] and references therein). The multistability can also exist in systems with small level of dissipation, or weakly dissipative systems, moreover, the number of coexisting attractors in such systems can be extremely high. For example, the existence of such a phenomenon was initially reported for the rotor map [4], and authors stated the existence of more than 100 coexisting attractors, while for the double rotor map the existence over than 3000 attractors was reported [5] at comparable level of dissipation. Later this phenomenon was discovered and investigated for a large amount of weakly dissipative systems, both for rather simple model equations [6] and for more realistic ones [7]. It turned out that the amount of attractors grows with the decrease of dissipation, but there are only few chaotic ones between them. Feudel and Grebodi [8] showed that this phenomenon is strongly concerned to the crises the attractor undergo with changes of parameters, both internal with change of dissipation parameter, and caused by the collision with the basin boundary with the change of nonlinearity parameter at constant level of dissipation. As a result, the chaotic attractor vanishes just after it appears from the period-doubling cascade. Changes also occur with basins of attraction and the basin boundaries [9]; first become rather small for the majority of the coexisting attractors, and the latter have rather complicated fractal-like structure in weakly dissipative systems.

The described scenario of an attractor death makes one to think about a possibility of coexistence of attractors with more complicated structure than periodic ones. In this context the idea of an external force on incommensurate frequency seems to be a productive one. Indeed, it is known, that such a forcing leads to complification of system dynamics: the limit cycles (fixed points in maps) become invariant tori (invariant curves in maps), and the strange nonchaotic attractor appears [10]. The multistable weakly dissipative system with such "quasiperiodical" forcing was considered recently [11], but in that work authors pay the main attention to the fact of existence of the strange nonchaotic attractor in such a system, while the problem of evolution of coexisting attractors remains practically untouched. In the present paper we are going to investigate these problems.

The structure of the paper is as follows: in Section 1 we introduce the system under investigation and consider the structure of bifurcation diagrams in order to reveal the dependence of the system dynamics on the nonlinearity and dissipation parameters. In Section 2 we consider the phase space of the system focusing on the basins of coexisting quasiperiodical attractors. Finally, we summarize the obtained results and discuss further perspectives of investigation in Section 3.

**1. The Ikeda map under quasiperiodical forcing**

Let us employ as a basic model for investigation the well-known Ikeda map

$$z_{n+1} = A + B z_n \exp(i(|z_n|^2 + \psi)). \qquad (1)$$

First it was introduced as a model describing the dynamics of light amplitude in the optical ring cavity [12], and later was shown to be applicable for the description of the nonlinear Duffing oscillator driven with external delta-pulses [13]. In the case of weak dissipation the Ikeda map demonstrates rather strong multistability, which was shown for its different modifications [6, 14]. It seems to be a good basic model for our purpose due to 2 reasons: (i) it does not have an attractor in the infinity, since every point on the phase plane belongs to some attractor basin; (ii) attractors originating from the primary and secondary resonances are well-distinguished on the bifurcation diagram in the case of weak dissipation [15].

Let us now introduce the quasiperiodical forcing into the system as a periodical modulation of the nonlinearity parameter $A$ with the frequency of this modulation incommensurate to the own frequency of the system. To determine it correctly, let us employ the model of pulse driven Duffing oscillator from [13]

$$\ddot{x} + \gamma \dot{x} + \omega_0^2 x + \beta x^3 = \sum C \delta(t - nT) \qquad (2)$$

and introduce there the modulation of the amplitude of the external forcing C

$$C(t) = C_0(1 + \varepsilon \cdot \sin(\omega t)).\qquad(3)$$

Reproducing calculations from [13] we get a map in the following form

$$z_{n+1} = A(1 + \varepsilon \sin(\Omega \cdot \psi \cdot n)) + Bz_n \exp(i|E_n|^2 + i\psi)\qquad(4)$$

where $\Omega = \omega/\omega_0$ is a parameter determining the ratio of external frequency to the own frequency of the system, and it has to be irrational. In all calculations parameter $\psi = 2\pi$ and $\Omega$ is chosen to be equal golden ratio, if it is not specially mentioned.

The parameter $\varepsilon$ determines here the amplitude of external forcing, and at $\varepsilon = 0$ the map (4) turns into the classical Ikeda map (1). Let us remind here the structure of its attractors and their dependence on parameters. In Figure 1 bifurcation diagrams at different values of dissipation are shown. Bifurcation diagrams are plotted for a number of initial conditions on one picture in order to visualize the dynamics of coexisting attractors [4]. Coexisting attractors here can be divided into two types [15]. Attractors of the first type appear earlier while increasing the parameter *B*, and are characterized by a regular dependence of their coordinates on the parameter and more or less long interval of their existence by the nonlinearity parameter *A*. On the other hand, attractors of the second type are characterized by more complex dependence on parameters and smaller parameter interval of their existence, and they appear at smaller values of dissipation. Attractors of the first type originate from the fixed points corresponding primary resonances, while the attractors of second type – resonances of higher orders.

Now let us start to increase the forcing amplitude $\varepsilon$. Figure 2 shows bifurcation diagrams for perturbed system (4) and nonperturbed system (1) plotted one on another at relatively sufficient dissipation value *B*=0.7 (panel a) and portraits of attractors of perturbed system at different values of nonlinearity parameter (panels b-d). One can see, that only few of the coexisting attractors survive in the perturbed system even at small perturbation amplitude, and none of them are of the second type. The reason is apparently the small scale of the basins of attraction, which do not leave place for the existence of the finite-sized torus attractor. The remained attractors demonstrate one torus-doubling bifurcation, and then the tori destruct after fractalization.

Decreasing of dissipation (*B*=0.9 in Fig. 3 a) leads to the increase of the number of the coexisting tori without sufficient qualitatively changes, but we can mention here, that the size of the attractor changes now non-monotonically during its interval of existence. The increase of the perturbation amplitude leads to further decrease of the number of the coexisting tori (Fig. 3 b, c). At *B*=0.99 (Fig. 4 a) the whole structure of the diagram becomes much more complicated. We can stress here that attractors of the first type, existing at small values of nonlinearity parameter *A*, remain small in size, while the size of secondary attractors grows, and they begin to overlap in the projection onto *x* (or *y*) axis visible in the bifurcation diagram. Moreover, at increase of $\varepsilon$ new type of attractors arise (Fig. 4 d) - the invariant curve surrounding the primary attractor. Such attractors also could not be detected via the bifurcation diagram due to the "shadowing" after projection on one of the coordinate axis. Here the problem of dividing of such attractors in order to count them and to divide their basins arises.

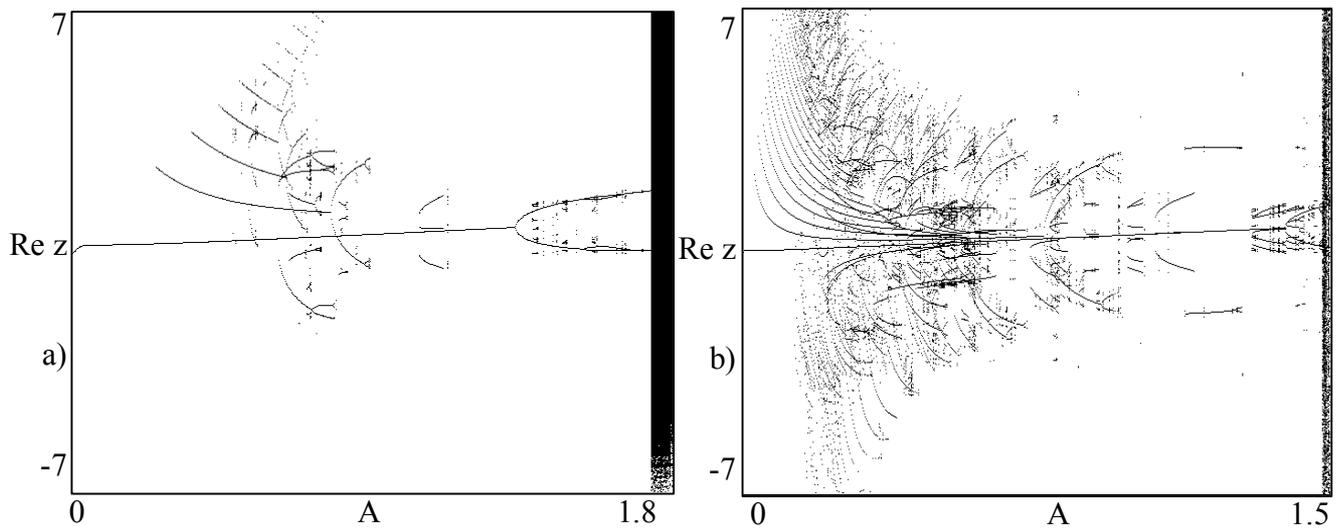

Fig. 1. Bifurcation diagrams of the Ikeda map (1) showing the multiple coexisting attractors at different levels of dissipation *B*: a) 0.9; b) 0.99.

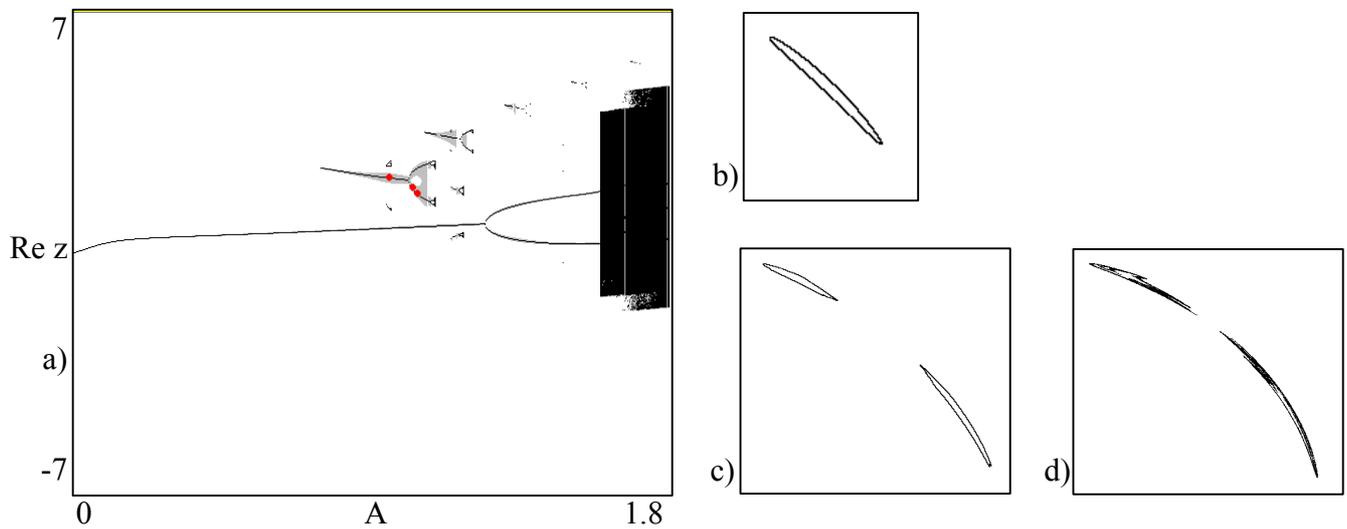

Fig. 2. Bifurcation diagrams for nonperturbed map (1) (black) and for the perturbed map (4) (light-grey) at *B*=0.7 (a) and portraits of attractors of the perturbed map (4) on the complex plane *z* (b-d). Parameter ε=0.01 for the map (4), values of parameter *A* corresponding the attractors in the panels (b)-(d) are marked in the panel (a) with the red points (from left to right respectively).

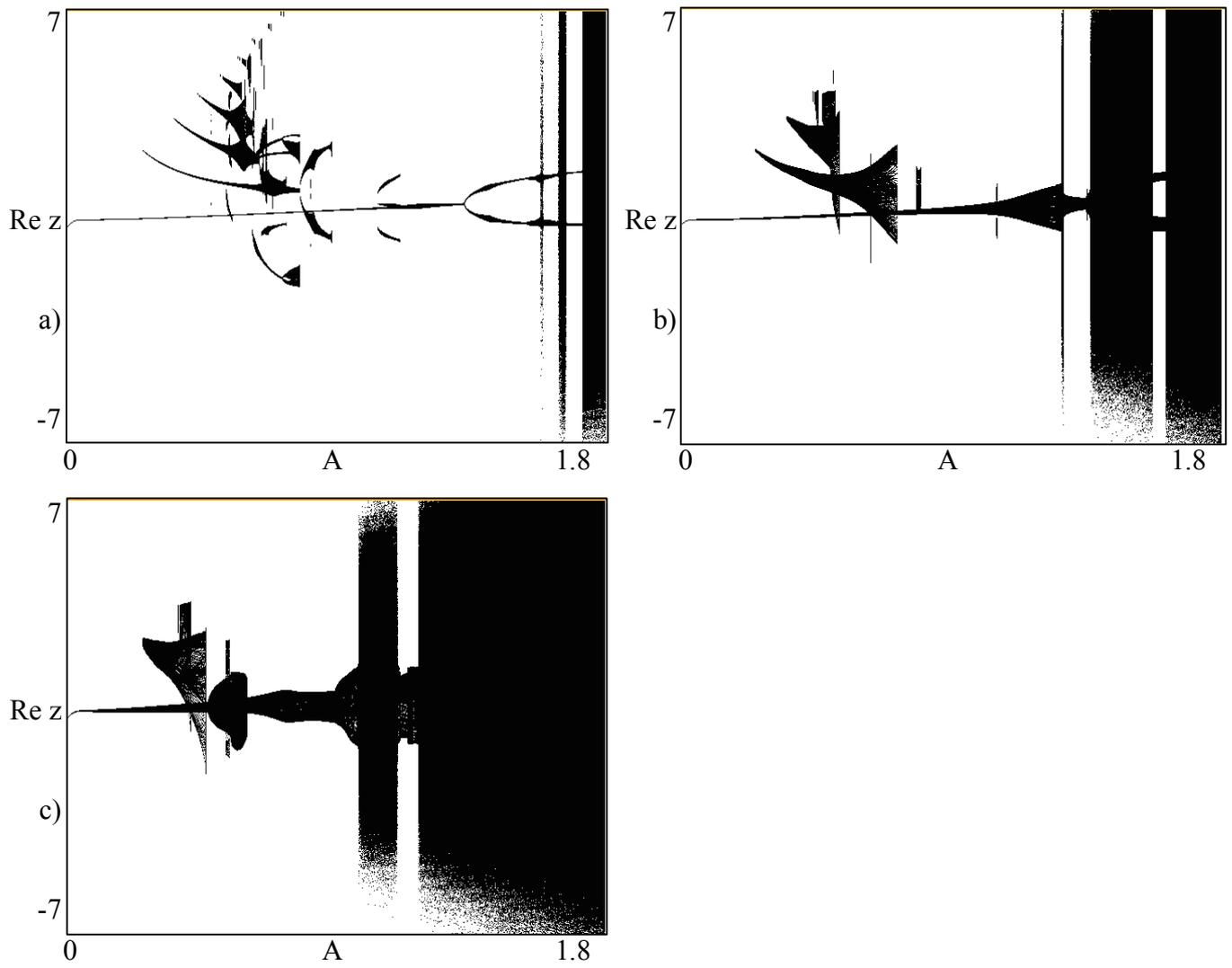

Fig. 3. Bifurcation diagrams for the perturbed map (4) at $B$=0.9 at different values of the external forcing amplitude ε: a) 0.01 b) 0.1 c) 0.5.

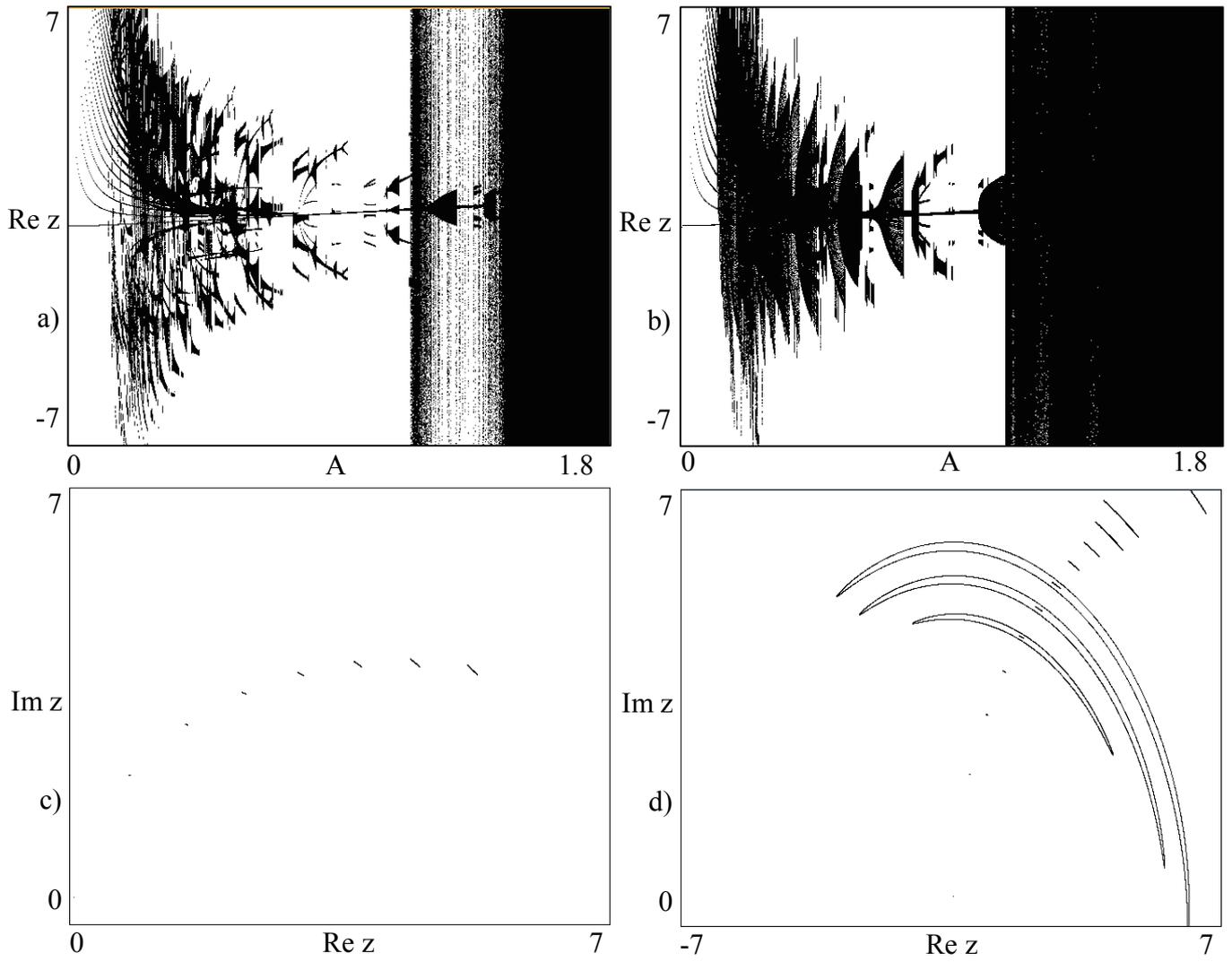

Fig. 4. Bifurcation diagrams for the perturbed map (4) at $B=0.99$ at different values of the external forcing amplitude $\varepsilon$: a) 0.01 b) 0.05, and portraits of attractors at $\varepsilon=0.05$, values of the nonlinearity parameter $A$: c) 0.084375 d) 0.140625.

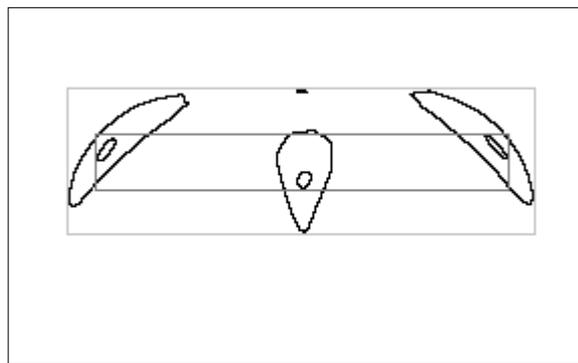

Fig. 5. The sketch illustrating the procedure of the attractor detection. The inner (dark-grey) and the outer (light-grey) rectangles determine two different invariant curves.

## 2. The structure of the phase space: the basins of attraction

In order to solve this problem we introduce the following method. For each attractor we determine four points with minimal and maximal values of the dynamical variables $(x_{min},y_1)$, $(x_{max},y_2)$, $(x_3,y_{min})$, $(x_4,y_{max})$. Obviously the attractor in this case is situated inside the rectangle with vertices $(x_{min},y_{min})$, $(x_{min},y_{max})$, $(x_{max},y_{max})$, $(x_{max},y_{min})$, and four points mentioned above are the points of tangencies of the torus and the rectangle (the sketch is shown in Figure 5). In general case such four points will determine the certain attractor, and each torus attractor could be identified by these four values.

Using described algorithm we have scanned the phase plane at certain parameter values in order to calculate the number of coexisting attractors and to investigate, how their number decreases with the growth of the amplitude ε. Also it is interesting to investigate the change of the relative size of basins of different attractors with the parameters. To evaluate this characteristics quantitatively, one can calculate the ratio of initial conditions belonging to the basin of the certain attractors to the total number of initial conditions. Of course, exact value of the relative size of the basin could be reached only for the infinite number of initial conditions, but one still can have some estimation of it. We obtained that the number of attractors and the relative size of their basins do not change sufficiently if the distance between neighbor points in the set of initial conditions is less than 0.1 by each coordinate. For our calculations we used sets of initial conditions with distance between points 0.025 by each coordinate.

Basins of attraction at the value of nonlinearity parameter A=0.6 and several values of the external forcing amplitude are shown in the Fig.6. It is well observed that the increase of the ε value leads to the increase of some basins and to the decrease and following vanishing of other ones. At ε=0.5 there exist two attractors, while the size of the main basin is about 98% of the phase plane in the considered area.

At *A*=0.2 (Fig. 7) the number of coexisting attractors is much more, but we obtain qualitatively the same behaviour when varying the ε parameter. All attractors except the main one vanish at ε≈1.5.

Figure 8 shows the dependence of number of coexisting attractors on the ε parameter at *A*=0.2 and 0.6. We can see that this dependence is a descending curve but in the area of small external forcing amplitude it is strongly non-monotonous. The reason seems to be as follows. The number and structure of coexisting attractors in the original map (1) depends strongly on the values of the nonlinearity parameter, and the modulation of the external forcing leads to the effective change of it.

## 3. Results and discussion

In this paper we have studied the model 2D map – the Ikeda system – under external forcing on incommensurate frequency. We have shown the coexistence of large number of stable invariant curves, and calculated how their number depends on the amplitude of the external forcing. At small external forcing amplitude quasiperiodical attractors originating from the primary resonances (first type) and resonances of higher periods (second type) coexist, while at increase of this amplitude attractors of the second type begin to grow in size and then destruct, while the new type of attractors appear: invariant curves surrounding the attractors of the first type. The number of coexisting attractors decrease with further increase of the external forcing amplitude, but non-monotonically.

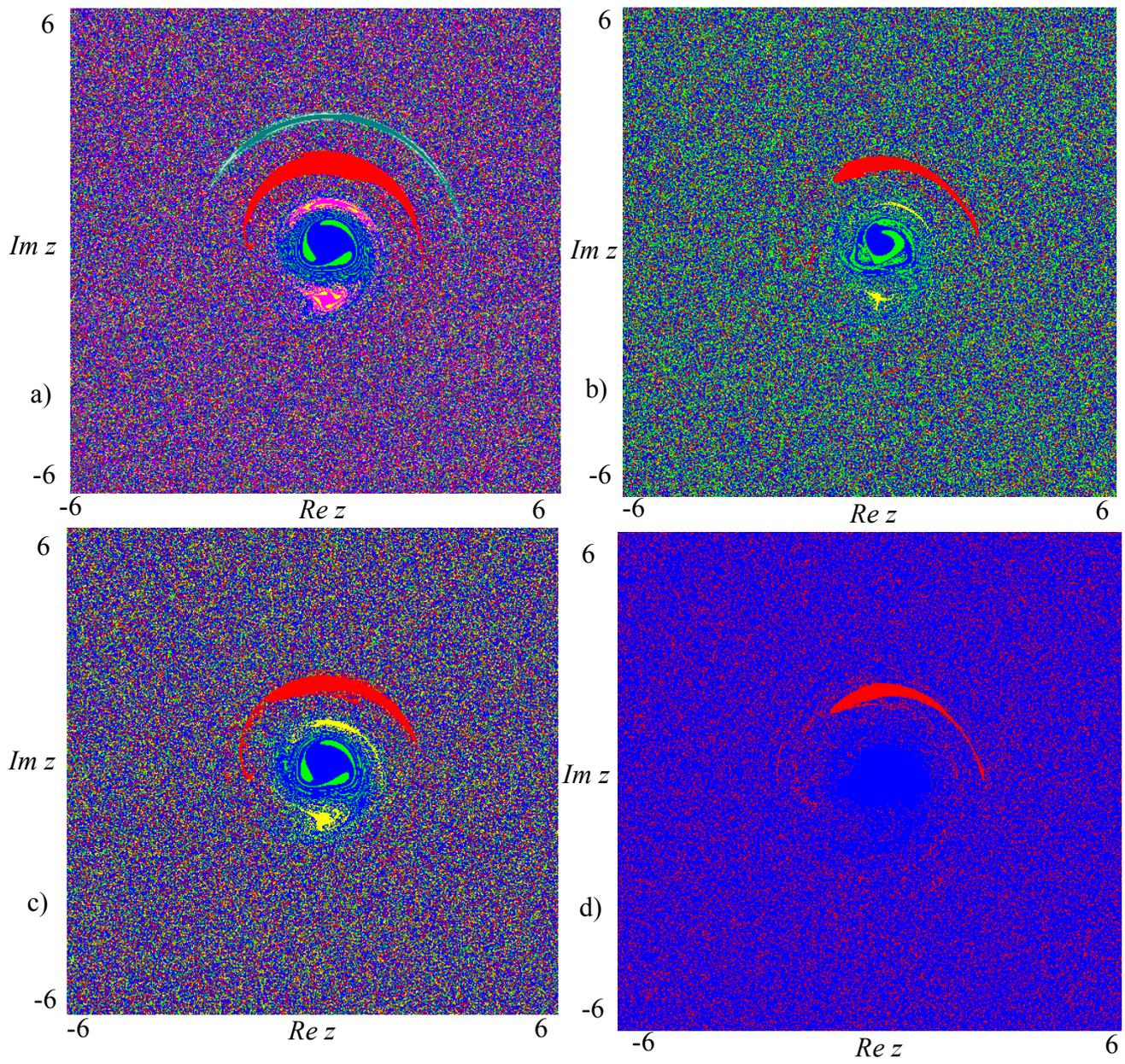

Fig 6. Basins of attraction at $A$=0.6, $B$=0.99, $\psi=\pi/2$ and different amplitudes of external forcing $\varepsilon$: a) 0, b) 0.1, c) 0.2, d) 0.3.

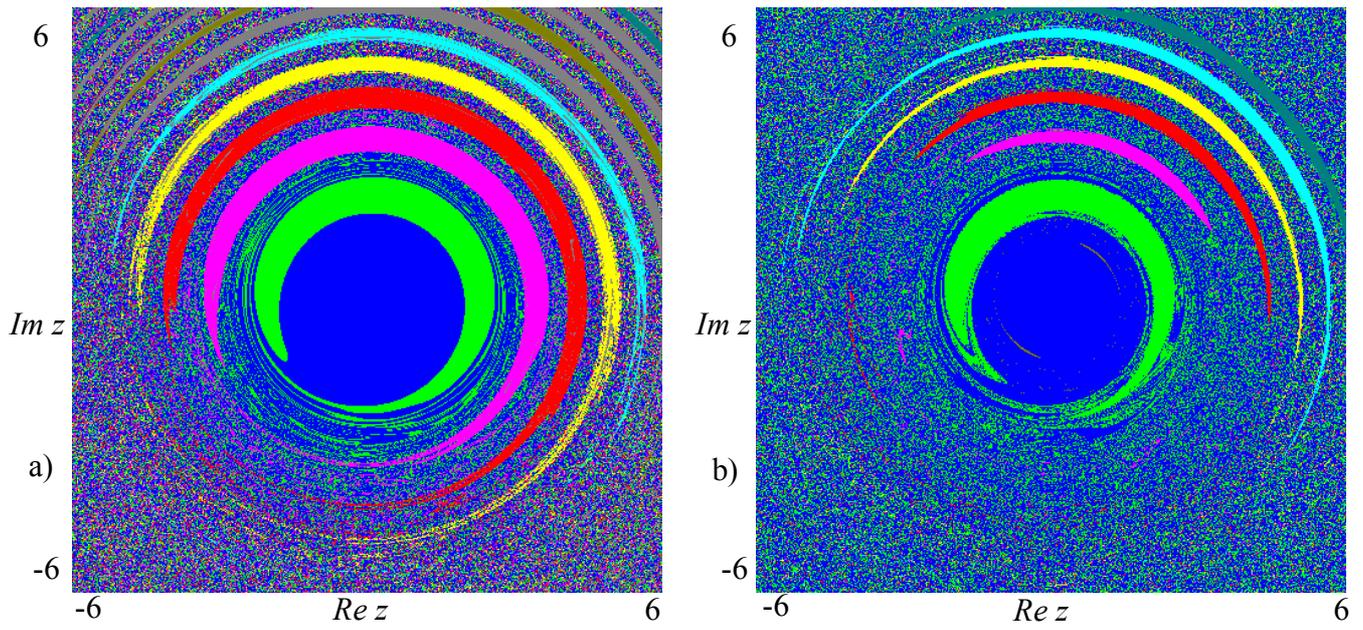

Fig 7. Basins of attraction at $A=0.2$, $B=0.99$, $\psi=\pi/2$ and different amplitudes of external forcing $\varepsilon$: a) 0, b) 0.4.

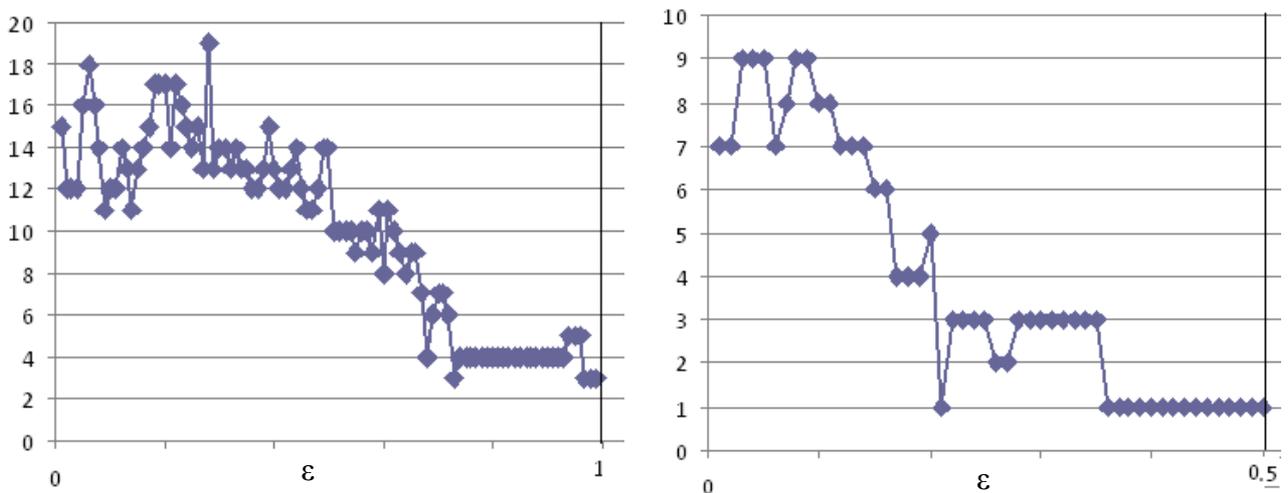

Fig. 8. Dependence of the number of coexisting attractors of the perturbed system (4) on the amplitude of the external forcing $\varepsilon$: $A=0.2$ (left) and $A=0.6$ (right).

We suppose that the reason is the modulation of the nonlinearity parameter of the original map by the external forcing.

Mechanisms of the attractor destruction and birth as well as the changes which the basin boundary undergo with the change of the forcing amplitude are not clear and require further investigations.

**Acknowledgements**
Authors would like to thank the the Russian Foundation for Basic Research for financial support through Project No. 14-02-31067.